\documentclass[twocolumn,showpacs,preprintnumbers,amsmath,amssymb,nofootinbib,pra,aps,10pt]{revtex4-1}
\usepackage{hyperref}
\usepackage{graphicx}
\usepackage{tikz}

\begin{document}

\title{Redshift drift of gravitational lensing}
\author{Oliver F. Piattella}
\email{oliver.piattella@pq.cnpq.br}
\author{Leonardo Giani}
\email{leonardo.giani@aluno.ufes.br}
\affiliation{Physics Department, Universidade Federal do Esp\'irito Santo, Vit\'oria, 29075-910, Brazil}

\date{\today}

\begin{abstract}
We investigate the effect of the redshift drift in strong gravitational lensing. The redshift drift produces a time variation of $i)$ the apparent position of a lensed source and $ii)$ the time delay among incoming signals from different images. We dub these effects as \textit{angular drift} and \textit{time delay drift}, respectively, and analyze their relevance in cosmology. 
\end{abstract}


\maketitle


\section{Introduction}

The redshift drift is the time variation of the redshift of a source due to the Hubble flow. It is a precious way of testing cosmologies in the \textit{redshift desert zone} (corresponding to redshifts in the interval $2 < z < 5$) because its measure is model-independent, i.e. it does not depend on \textit{a priori} assumptions on the matter content but it only relies on the background Friedmann-Lema\^itre-Robertson-Walker (FLRW) geometry \cite{1962ApJ...136..319S, Loeb:1998bu, Corasaniti:2007bg, Balbi:2007fx, Uzan:2007tc, Uzan:2008qp, Quartin:2009xr, Jain:2009bm, Quercellini:2010zr, Moraes:2011vq, Stebbins:2012vw, Neben:2012wc, Balcerzak:2012bv, Martinelli:2012vq, Darling:2012jf, Li:2013oba, Yuan:2013wpa, Zhang:2013zyn, Kim:2014uha, Geng:2014hoa, Geng:2014ypa, Zhang:2014bwt, Geng:2015ara, Geng:2015hen, Koksbang:2015ctu, Guo:2015gpa, Martins:2016bbi, He:2016rvp, Melia:2016bnb}. The importance of the redshift drift was appreciated for the first time by Sandage in 1962 \cite{1962ApJ...136..319S}, and its application for the measurement of the temporal shift of quasar Lyman-$\alpha$ absorption lines was proposed by Loeb in 1998 \cite{Loeb:1998bu} in what is now known as \textit{Sandage-Loeb test} \cite{Corasaniti:2007bg}. 

At his time, Sandage's conclusion was that unfortunately astronomers would have needed about $10^7$ years in order to detect an appreciable and useful effect. This made the observation of the redshift drift unfeasible, in the least. On the other hand, impressive progress has been made in spectroscopy since 1962 and the measurement of the redshift drift seems to be not that hopeless, requiring observation times of the order of 10 years, according to Ref.~\cite{Martins:2016bbi}, rather than $10^7$ years. We refer, for example, to the ELT-HIRES experiment.\footnote{\url{http://www.hires-eelt.org}}

This exciting perspective has recently raised some interest in the redshift drift. In particular, the authors of Ref.~\cite{Martins:2016bbi} calculate the second time derivative of the redshift, which is related to the \textit{jerk} parameter, and the author of Ref.~\cite{Melia:2016bnb} proposes the measurement of the redshift drift as a definitive test for $R_h = ct$ cosmology, for which the redshift drift is vanishing.

In this paper we analyze how the redshift drift manifests itself in strong gravitational lensing, providing two extra ways, beyond spectroscopy, of determining the value of the Hubble parameter at the lens redshift. In particular, these are: $i)$ the drift of the apparent angular position of a lensed source and $ii)$ the drift of the time delay among incoming signals from multiple images of a lensed source. We consider the simple case of a point-like lens and show that both the drifts are proportional to $H_0 - H_L/(1 + z_L)$, where $z_L$ is the lens redshift and $H_L$ is the Hubble parameter evaluated at $z_L$. The measure of such drifts, especially the angular one, seems to be unfeasible at present time, but it might be an interesting observational challenge for the future.  

The paper is divided as follows. We present the basic equations governing the redshift drift in Sec.~\ref{Sec:RedDr}. In Secs.~\ref{Sec:Angledrfit} and \ref{Sec:TimeDeldrift} we calculate the drifts of the image angle and of the time delay. Finally, Sec.~\ref{Sec:DiscandConcl} is devoted to our discussion and conclusions. We use $c = 1$ units throughout the paper.


\section{The redshift drift}\label{Sec:RedDr}

From the FLRW metric
\begin{equation}\label{FLRWmetric}
	ds^2 = -dt^2 + a(t)^2\gamma_{ij}dx^idx^j\;,
\end{equation}
with
\begin{equation}
	\gamma_{ij}dx^idx^j = \frac{dr^2}{1 - Kr^2} + r^2d\Omega^2\;,
\end{equation}
where $K$ is the spatial curvature, one straightforwardly obtains the result that a photon is redshifted, and the redshift is given in terms of the scale factor as follows:
\begin{equation}
	1 + z = \frac{a_0}{a_e}\;,
\end{equation}
where $a_0$ is the scale factor evaluated at present time (which is the time of observation) and $a_e$ is the scale factor evaluated at the emission time.

The derivative of the redshift with respect to the observation time $t_0$ is the following:
\begin{equation}\label{reddriftformula}
	\frac{dz}{dt_0} = \frac{1}{a_e}\frac{da_0}{dt_0} - \frac{a_0}{a_e^2}\frac{da_e}{dt_0} = \frac{1}{a_e}\frac{da_0}{dt_0} - \frac{a_0}{a_e^2}\frac{da_e}{dt_e}\frac{dt_e}{dt_0}\;.
\end{equation}
From the FLRW metric \eqref{FLRWmetric} it is not difficult to show that:
\begin{equation}
	\frac{dt_e}{dt_0} = \frac{a_e}{a_0}\;,
\end{equation}
and therefore, the redshift drift formula \eqref{reddriftformula} becomes:
\begin{equation}\label{reddriftformula2}
	\frac{dz}{dt_0} = \frac{1}{a_e}\frac{da_0}{dt_0} - \frac{1}{a_e}\frac{da_e}{dt_e} = (1 + z)H_0 - H(z)\;,
\end{equation}
where the Hubble constant $H_0$ and the Hubble parameter $H(z)$ have appeared. Measuring $dz/dt_0$ would allow to determine $H(z)$, thereby providing precious informations on the energy content of the universe.

In the next sections we shall obtain formulae similar to the one in Eq.~\eqref{reddriftformula2}, but in the context of strong gravitational lensing.


\section{Angular drift}\label{Sec:Angledrfit}

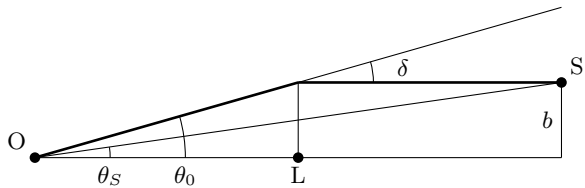
\begin{figure}[h!]
\centering
	\begin{tikzpicture}
	\draw (0,0) -- (7,0);
	\draw (7,0) -- (7,1);
	\draw [line width=1pt] (3.5,1) -- (7,1);
	\draw [line width=1pt] (0,0) -- (3.5,1);
	\draw (3.5,0) -- (3.5,1);
	\fill (0,0) circle (2pt) node[above left] {O};
	\fill (3.5,0) circle (2pt) node[below] {L};
	\fill (7,1) circle (2pt) node[above right] {S};
	\draw (0,0) -- (7, 2);
	\draw (0,0) -- (7, 1);
	\draw (4.5,1) arc (0:16:1);
	\draw (4.7,1.2) node[right] {$\delta$};
	\draw (1,0) arc (0:8:1);
	\draw (2,0) arc (0:16:2);
    \draw (1,0) node[below] {$\theta_S$};
    \draw (2,0) node[below] {$\theta_0$};
    \draw (7,0.5) node[left] {$b$};
\end{tikzpicture}
\caption{Scheme for gravitational lensing. The deflection angle is $\delta = 4GM/b$, where $b$ is the impact parameter.}
\label{Fig:lensfig}
\end{figure}

Let's consider the lens equation in the thin-lens approximation \cite{Weinberg:2008zzc}:
\begin{equation}\label{lenseq}
	\theta_0(\theta_0 - \theta_S) = \frac{4GM}{d_L}\frac{d_{LS}}{d_S}\;,
\end{equation}
where $\theta_0$ is the angular position of the image, $\theta_S$ is the angular position of the source, $M$ is the lens mass and with $d$ we denote angular-diameter distances. In Fig.~\ref{Fig:lensfig} we display the lensing scheme.

In particular, the angular-diameter distance to the source is written as:
\begin{equation}\label{dS}
	d_S \equiv a_S\chi_S = \frac{1}{1 + z_S}\int_0^{z_S}\frac{dz'}{H(z')}\;,
\end{equation}
with $\chi_S$ the comoving distance to the source, $a_S$ is the scale factor at the emission time and $z_S = 1/a_S - 1$ is the redshift of the source. 

Similarly, the angular-diameter distance between the lens and the source is the following:
\begin{equation}\label{dLS}
	d_{LS} \equiv a_S(\chi_S - \chi_L) = \frac{1}{1 + z_S}\int_{z_L}^{z_S}\frac{dz'}{H(z')}\;,
\end{equation}
and the angular-diameter distance to the lens is the following:
\begin{equation}\label{dL}
	d_{L} \equiv a_L\chi_L = \frac{1}{1 + z_L}\int_{0}^{z_L}\frac{dz'}{H(z')}\;,
\end{equation}
with $\chi_L$ being the comoving distance to the lens and $z_L$ the redshift of the lens. With these definitions, let's rewrite the lens equation as follows:
\begin{equation}\label{lenseq2}
	\theta_0(\theta_0 - \theta_S) \equiv \theta_E^2 = 4GM(1 + z_L)\left(\frac{1}{\chi_L} - \frac{1}{\chi_S}\right)\;,
\end{equation}
where $\theta_E$ is the Einstein ring radius, if $\theta_S = 0$. From the first part of the the above equation we have the following two solutions for $\theta_0$:
\begin{equation}
\theta_\pm = \frac{\theta_S}{2} \pm \sqrt{\frac{\theta_S^2}{4} + \theta_E^2}\;,
\end{equation}
i.e. two solutions $\theta_\pm$ for the images, given the high symmetry of a point-like lens. Combining the two solutions, we get:
\begin{equation}\label{thetapmrel}
	\theta_S = \theta_+ + \theta_-\;, \qquad \theta_E^2 = -\theta_+\theta_-\;,
\end{equation}
where notice that $\theta_-$ is negative. Notice also that $\theta_S$ is time-independent because the Hubble flow moves lens and source radially with respect to the observer (us), thereby leaving $\theta_S$ unchanged, i.e. $d\theta_S/dt_0 = 0$ and thus $d\theta_+/dt_0 = -d\theta_-/dt_0$. The time derivatives of the observed angles are thus directly related to the time derivative of $\theta_E$ as follows, using Eq.~\eqref{thetapmrel}:
\begin{equation}
\frac{2d\theta_E}{dt_0} = -\theta_-\frac{d\theta_+}{dt_0} - \theta_+\frac{d\theta_-}{dt_0} = (\theta_+ - \theta_-)\frac{d\theta_+}{dt_0}\;.
\end{equation}
We can eliminate $\theta_-$ or $\theta_+$ from the above equation, by using again Eq.~\eqref{thetapmrel}, and obtain:
\begin{equation}\label{logdtheta0thetaE}
\frac{1}{\theta_0}\frac{d\theta_0}{dt_0} = \frac{2\theta_E^2}{\theta_0^2 + \theta_E^2}\frac{1}{\theta_E}\frac{d\theta_E}{dt_0}\;.
\end{equation}
Having established this correspondence between the two derivatives, we now proceed with the calculations of the time drift of $\theta_E$. Since the comoving distance is time-independent, the derivative with respect to the observer time of the Einstein radius is the following, from Eq.~\eqref{lenseq2}:
\begin{eqnarray}
	2\theta_E\frac{d\theta_E}{dt_0} = 4GM\frac{dz_L}{dt_0}\left(\frac{1}{\chi_L} - \frac{1}{\chi_S}\right)\;.
\end{eqnarray}
Using Eq.~\eqref{reddriftformula2}, one can write the relative variation of the angle $\theta_E$ in the following form:
\begin{equation}\label{angdriftform}
	\frac{2}{\theta_E}\frac{d\theta_E}{dt_0} = H_0 - \frac{H_L}{1 + z_L}\;.
\end{equation}
Not unexpectedly, this formula is very similar to the one in Eq.~\eqref{reddriftformula2}. If $H_L > H_0(1 + z_L)$, we expect the Einstein radius to shrink. This is the case for a matter-dominated universe, for example.


\section{Time delay drift}\label{Sec:TimeDeldrift}

The time-delay is usually divided into a geometric contribution and a potential one \cite{Weinberg:2008zzc}. The former is due to the bending of the trajectory of the photon, whereas the latter is due to the motion into the lens gravitational field. 

The geometric time delay for a single image is given by:
\begin{equation}\label{geometrictimedelay}
	\Delta t_{\rm geom} = \frac{(1 + z_L)d_Sd_L(\theta_0 - \theta_S)^2}{2d_{LS}}\;.
\end{equation}
This formula, being purely geometric, does not depend on the particular density profile of the lens. For a point-like lens, using Eq.~\eqref{lenseq2}, we can cast it as:
\begin{equation}
	\Delta t_{\rm geom} = \frac{(1 + z_L)^2(4GM)^2}{2\theta_0^2}\left(\frac{1}{\chi_L} - \frac{1}{\chi_S}\right)\;.
\end{equation}
The interesting quantity is of course the difference in the geometric time delays of the two images, i.e.
\begin{equation}
	\Delta_{\rm geom} \equiv \Delta t_{\rm geom}(\theta_+) - \Delta t_{\rm geom}(\theta_-)\;,
\end{equation}
which, using Eqs.~\eqref{lenseq} and \eqref{geometrictimedelay}, can be written as:
\begin{equation}
	\Delta_{\rm geom} = 2GM(1 + z_L)\frac{\theta_+^2 - \theta_-^2}{\theta_+\theta_-}\;.
\end{equation}
Its logarithmic time-derivative is, using Eqs.~\eqref{reddriftformula2}, \eqref{logdtheta0thetaE} and \eqref{angdriftform}:
\begin{equation}
	\frac{1}{2\Delta_{\rm geom}}\frac{d\Delta_{\rm geom}}{dt_0} = \left(H_0 - \frac{H_L}{1 + z_L}\right)\frac{\theta_E^2}{(\theta_+ - \theta_-)^2}\;.
\end{equation}
Again, between parenthesis we have the same factor as in Eq.~\eqref{angdriftform}. It is multiplied by a combination of angles which is always less than 1/4, being however this value correspondent to the case of an Einstein ring, for which $\Delta_{\rm geom} = 0$.

The potential time delay is a local effect related to the lens gravitational potential. It can be written as:
\begin{equation}
	\Delta t_{\rm pot} = (1 + z_L)f(b)\;,
\end{equation}
where $f(b)$ is a function which depends on integrals of the gravitational potential and $b$ is the impact parameter \cite{Weinberg:2008zzc}. The explicit expression of $f$ is quite cumbersome, so we do not report it here. What is important is that for a point-like lens, the relative potential time delay can be expressed as follows:
\begin{equation}
	\Delta_{\rm pot} \equiv \Delta t_{\rm pot}(\theta_+) - \Delta t_{\rm pot}(\theta_-) = 2GM(1 + z_L)\ln\frac{-\theta_-}{\theta_+}\;,
\end{equation}
where the minus sign in the logarithm comes from the fact that $\theta_- < 0$. By differentiation we straightforwardly obtain:
\begin{eqnarray}
	\frac{1}{\Delta_{\rm pot}}\frac{d\Delta_{\rm pot}}{dt_0} = H_0 - \frac{H_L}{1 + z_L}\nonumber\\ + \frac{1}{\ln\frac{-\theta_-}{\theta_+}}\left(\frac{1}{\theta_-}\frac{d\theta_-}{dt_0} - \frac{1}{\theta_+}\frac{d\theta_+}{dt_0}\right)\;.
\end{eqnarray}
Using Eqs.~\eqref{logdtheta0thetaE} and \eqref{angdriftform}, we can cast the above relation as follows:
\begin{eqnarray}
	\frac{1}{\Delta_{\rm pot}}\frac{d\Delta_{\rm pot}}{dt_0} = \left(H_0 - \frac{H_L}{1 + z_L}\right)\nonumber\\\left(1 - \frac{1}{\ln\frac{-\theta_-}{\theta_+}}\frac{\theta_- + \theta_+}{\theta_- - \theta_+}\right)\;.
\end{eqnarray}
The logarithm is vanishing for $\theta_+ = -\theta_-$, which is the case of an Einstein ring and therefore no time delay exists. 

It must be said that one cannot disentangle the geometric from the potential delay and both are of order $GM$. On the other hand, as we have explicitly calculated above, the two time delays drift in different manners. Therefore, it might be possible to distinguish one from the other if the drift was dominated by one of the two, which is not the case for the point-like lens that we have considered here.

The total time delay between the two images is thus $\Delta \equiv \Delta_{\rm geom} + \Delta_{\rm pot}$ and its drift:
\begin{eqnarray}\label{totaltimedelaydriftform}
	\frac{1}{\Delta}\frac{d\Delta}{dt_0} = \left(H_0 - \frac{H_L}{1 + z_L}\right)\frac{\ln\frac{-\theta_-}{\theta_+} + \frac{\theta_- + \theta_+}{\theta_- - \theta_+}}{\ln\frac{-\theta_-}{\theta_+} + \frac{\theta_+^2 - \theta_-^2}{\theta_+\theta_-}}\;,
\end{eqnarray}
where the angular contribution on the right hand side is of order unity.


\section{Discussion and conclusions}\label{Sec:DiscandConcl}

The Hubble flow should manifest itself in a time drift of the redshift of a source. In this paper, we have explored other two possible indications of the Hubble flow, in the context of strong gravitational lensing. These are the time drift of the apparent angular position of a lensed source, or \textit{angular drift}, and the drift in the delay of the arrival times of the signals coming from two different images, or \textit{time delay drift}.  

As we have computed for the simple case of a point-like lens, the entity of these variations is
\begin{equation}
	\frac{1}{\theta_E}\frac{d\theta_E}{dt_0} \sim \frac{1}{\Delta}\frac{d\Delta}{dt_0} \sim H_0 - \frac{H_L}{1 + z_L}\;,
\end{equation} 
cf. Eqs.~\eqref{angdriftform} and \eqref{totaltimedelaydriftform}. Therefore, measuring an angular drift or a time delay drift would make it possible to determine the value of the Hubble parameter evaluated at the lens redshift, i.e. $H_L$.

As we mentioned in the Introduction, the measurement of the redshift drift through spectroscopy seems to require observation times of the order of 10 years, so what about the angular and time delay drift? Let's make a crude estimate by using data that can be found in Ref.~\cite{Kochanek:2003pi}. Consider the object QSO0957+561, at redshift $z_S = 1.41$ which displays two images separated by $6.1''$ as it is lensed by a cluster at $z_L = 0.36$. The time-delay is $417 \pm 3$ days (with 95\% confidence level). Since $H_0 \approx 10^{-18}$ s$^{-1}$, then the angular drift would be of 
\begin{equation}
	\frac{d\theta_E}{dt_0} \sim 10^{-10} \quad \mbox{arc seconds per year}\;,
\end{equation}
which is certainly something very difficult to measure since, in order to arrive to a signal of $0.1$ arc seconds (of the same order of the precision of the measurement) one should need $10^9$ years.

For the time delay drift one has:
\begin{equation}
	\frac{d\Delta}{dt_0} \sim 10^{-3} \quad \mbox{seconds per year}\;,
\end{equation}
so that, in order to accumulate a drift of the order of the day, one should need $10^8$ years. These huge numbers do not make the observation of the drifts viable, but still they might be an interesting observational challenge for the future for decade-long monitoring programs of lensed quasars, such as COSMOGRAIL.\footnote{\url{http://www.cosmograil.org/}.}


\acknowledgments{The author thanks CNPq (Brazil) and FAPES (Brazil) for partial financial support.}


\bibliographystyle{unsrt}
\bibliography{Lensing}

\end{document}